\documentclass[prl,twocolumn,superscriptaddress,amsmath,amssymb]{revtex4}
\usepackage{graphicx}
\usepackage{indentfirst}
\usepackage{psfrag}
\usepackage{epsfig}
\usepackage{amsmath}
\usepackage{amssymb}
\usepackage{bm}
\usepackage{mathrsfs}
\newcommand{\be}{\begin{eqnarray}}
\newcommand{\ee}{\end{eqnarray}}

\begin{document}

\title{The  optical torque on small bi-isotropic particles} 
\author{Manuel Nieto-Vesperinas }
\affiliation{Instituto de Ciencia de Materiales de Madrid, Consejo Superior de
Investigaciones Cient\'{i}ficas\\
 Campus de Cantoblanco, Madrid 28049, Spain.\\ www.icmm.csic.es/mnv; 
mnieto@icmm.csic.es }
\begin{abstract}Most previous theoretical studies on the optical torque exerted by light on dipolar particles are incomplete. Here we establish the equations for the time-averaged optical torque on dipolar bi-isotropic particles.  Due to the interference of scattered fields, it has a term additional to that commonly employed in theory and experiments. Its consequences for conservation of energy, angular momentum, and effects like negative torques, are discussed.
\end{abstract}



\maketitle


In recent times, and with the steady  improvement of particle manipulation techniques, the torque exerted by light on particles is gaining increasing attention. However, in spite of the last advances attained on the  theory and experiment interpretation  of the optical force  (OF) on small particles \cite{mazilu,chaumet,arias,albala,berry,chaumet3,MNV2010},    there is no yet a parallel one on the optical torque (OT).  Most studies \cite{laporta,dunlop} have no quantitative assesment with theory or, if theoretical, involve a static formulation \cite{chaumet3,chaumet2, dogariu, cana} and thus  are  incomplete and hence incorrect: they do  not comply neither with the conservation of angular momentum nor with that of energy.  

We know few exceptions: reports involving Mie's calculations for larger particles or sets of spheres \cite{mars,chen1}. However, the latter convey computational procedures that, although yielding rigorous exact resuts, do not completely allow to see through analytic calculations the details of the different involved  physical processes. These, like in the case of optical forces, are clearly illustrated in the equations for the torque on a  dipolar object that, being a canonical system, is of great utility both for forthcoming incresingly controlled experiments and  interpretation purposes. This has already been extensively seen in the study of photonic forces.

In this letter we establish the theory of the OT on dipolar particles. By "dipolar" we mean those in the wide sense,  i.e.  whose polarizabiliities are fully given by the first electric and magnetic partial waves \cite{MNV2010, Nieto2011}, (e.g. Mie coefficients if they are spheres).  We put forward  that the time-averaged OT $<\bf \Gamma>$ on a dipolar particle contains a component additional  to that $<\bf \Gamma_0>$ (named {\it  intrinsic torque}) commonly derived from static equations and that, for reasons later discussed, coincides with what we shall name {\it extinction  optical torque} (EOT). This additional  component is the {\it recoil optical torque} (ROT) due to interference of  fields scattered by the particle, and is essential to fulfill the conservation of angular momentum and energy. It is curious, on the other hand, that this ROT implicitely appears as a well-known textbook result \cite{jackson} for the rate of angular momentum  radiated by a dipole, even though apparently without being noticed by many modern research works. However, the ROT is essential (as noticed in \cite{chen1} for sets of large particles) to predict new OT effects, such as e.g. the appearence of {\it negative optical torques} (NOT). A phenomenon which keeps an analogy with pulling forces \cite{chen, novits,dogariu1}, and that it is much more common than the latter, nor requiring a special combination of incident beam structure and object composition and morphology.

We   consider monochromatic light fields, so that the electric and magnetic vectors are:  ${\bf \cal E}({\bf r},t)$ and  ${\bf \cal B}({\bf r},t)$:  ${\bf \cal E}({\bf r},t)=\Re [{\bf E}({\bf r}) \exp(-i\omega t)]$ and  ${\bf \cal B}({\bf r},t)=\Re [{\bf B}({\bf r}) \exp(-i \omega t)]$. $\Re$ denotes real part.  We will address time-averages of electromagnetic quantities, including  the electric and magnetic  energy densities,  respectively:  $<w_e>=(\epsilon /16 \pi) |{\bf E}|^2$ and  $<w_m>=(1/16 \pi \mu) |{\bf B}|^2$,  and the time-averaged Poynting vector:  $<{\bf S}>=(c/8\pi \mu)\Re ({\bf E} \times{\bf B}^{*})$.

Let a particle of index $n_p$ be embedded in a medium with dielectric permittivity $\epsilon$ and magnetic permeability $\mu$, illuminated by an electromagnetic field whose electric and
magnetic vectors are: ${\bf E}^{(i)}$, ${\bf B}^{(i)}$. The optical torque ${\bf \Gamma}=d{\bf J}/dt$, exerted by this field on the object, comes from the conservation of the angular momentum ${\bf J}={\bf r} \times {\bf S}$ , ${\bf r}$ denoting a position vector of space. On time averaging, such conservation law:  $<{\bf \Gamma}>=<d{\bf J}_{mech}/dt> +<d{\bf J}_{field}/dt >=\int_{s}{\bf r} \times <{\bf T}>\, dS$ \cite{jackson} leads to the torque time-average:
\be
<{\bf \Gamma}>= \frac{1}{8 \pi} \Re \int_{S} {\bf r}\times [\epsilon ( {\bf E}  \cdot
{\bf s}) {\bf E}^{*} \nonumber \\
 + \mu^{-1} ( {\bf B} \cdot {\bf s}) {\bf
B}^{*} - \frac{1}{2} (\epsilon | {\bf E}|^2+ \mu^{-1} | {\bf
B}|^2) {\bf s} ] d S.  \,\,({\bf r}=r{\bf s}). \,\,\,\,\,\label{torque1}
\ee
The angular momentum $<{\bf J}>$ has been splitted into the mechanical one: $<{\bf J}_{mech}>$ and that of the wave: $<{\bf J}_{field}>$, ($<d{\bf J}_{field}/dt >=0$ for these fields).  ${\bf T}$ is Maxwell's stress tensor.  $dS$ denotes the element of any surface $S$ (which will be considered a sphere for calculation purposes) that encloses the particle, and $\bf s$ is its local unit outward normal.   ${\bf r}=r {\bf s}$  is a point of $S$.

The wavefields in Eq.(\ref{torque1}) are total fields: the
sum of those incident and scattered, denoted by the $(i)$ and $(s)$-superscripts: ${\bf
E}^{(i)}+{\bf E}^{(s)}$,  ${\bf B}^{(i)}+{\bf B}^{(s)}$. Obviously the last two terms of the integrand of  (\ref{torque1}) do not contribute to $<{\bf \Gamma}>$.

Let the particle be magnetodielectric  and bi-isotropic \cite{kong}, dipolar in the wide sense,  for example if  a sphere \cite{MNV2010, Nieto2011}, its  magnetodielectric response is characterized by its electric,  magnetic, and magnetoelectric  polarizabilities :  $\alpha_{e}=i\frac{3}{2k^{3}}a_{1}$,  $\alpha_{m}=i\frac{3}{2k^{3}}b_{1}$,  $\alpha_{em}=i\frac{3}{2k^{3}}c_{1}$,  $\alpha_{me}=i\frac{3}{2k^{3}}d_{1}$.  $a_{1}$, $b_{1}$ and  $c_{1}$ standing for the electric, magnetic, and magnetoelectric \cite{chanlat} first Mie coefficients, respectively. Then the electric and magnetic dipole moments, $ {\bf p}$ and  ${\bf m}$, induced  on the particle by the illuminating field,  are
\be
{\bf p}=\alpha_{e} {\bf E}^{(i)}+\alpha_{em}{\bf  B}^{(i)}, \,\,\,\,\,
{\bf m}=\alpha_{me}{\bf E}^{(i)}+\alpha_{m}{\bf B}^{(i)}. \label{consti}
\ee

 First consider an incident plane wave, circularly polarized (CP), (cf. Fig.1):   ${\bf E}^{(i)}={\bf e}_{i} e^{ik({\bf s}_{i}\cdot {\bf r})}$,  ${\bf B}^{(i)}={\bf b}_{i} e^{ik({\bf s}_{i}\cdot {\bf r})}$. The scattered fields at a point $P$: $r{\bf s}$  are
\be
 {\bf E}^{(s)}=\frac{1}{\epsilon}[3{\bf s}({\bf s}\cdot{\bf p})-{\bf p}](\frac{1}{r^3}-\frac{ik}{r^2}) \exp(ikr) \nonumber \\
-i \sqrt{\frac{\mu}{\epsilon}}k({\bf s}\times{\bf m})\frac{e^{ikr}}{r^2}+{\bf e}({\bf s}_{xy}) \frac{\exp(ikr)}{r}, \,\,\, \,\,\,\,\,\, \label{es}
\ee
\be
 {\bf B}^{(s)}=\mu[3{\bf s}({\bf s}\cdot{\bf m})-{\bf m}](\frac{1}{r^3}-\frac{ik}{r^2}) \exp(ikr) \nonumber \\
+i \sqrt{\frac{\mu}{\epsilon}}k({\bf s}\times{\bf p})\frac{e^{ikr}}{r^2}+{\bf b}({\bf s}_{xy}) \frac{\exp(ikr)}{r } , \,\,\, \,\,\,\,\,\, \label{bs}   .
\ee
With the far-zone scattering amplitudes being
\begin{equation}
{\bf e}({\bf s}_{xy})=k^2  [\frac{1}{\epsilon}({\bf s} \times {\bf
p})\times {\bf s}- \sqrt{\frac{\mu}{\epsilon}}({\bf s}\times
{\bf m})], \label{dipe}
\end{equation}
\begin{equation}
{\bf b}({\bf s}_{xy})=k^2 [\mu({\bf s} \times {\bf m})\times
{\bf s}+\sqrt{\frac{\mu}{\epsilon}} ({\bf s}\times {\bf p})]. \label{dipm}
\end{equation}
${\bf s}=({\bf s}_{xy}, s_z)$,  ${\bf b}_{i}=n {\bf s}_{i} \times  {\bf e}_{i}$, ${\bf e}_{i} \cdot {\bf s}_{i}= {\bf b}_{i} \cdot {\bf s}_{i}=0$;  ${\bf b}=n {\bf s} \times  {\bf e}$, ${\bf e} \cdot {\bf s}= {\bf b} \cdot {\bf s}=0$. The rest of terms of (\ref{es}) and (\ref{bs}) are the fields in the near and intermediate zones \cite{jackson}.

\begin{figure}[htbp]
\centerline{\includegraphics[width=.8\columnwidth]{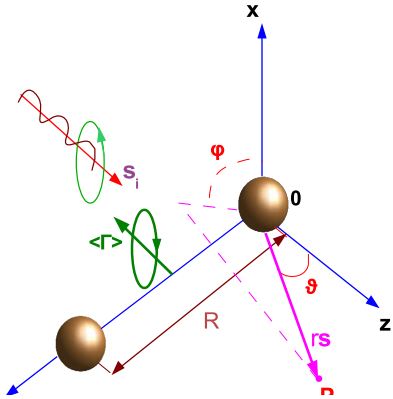}}
\caption{(Color online). Scattering geometry illustrating a negative optical torque of a CP plane wave on two particles. }
\end{figure}



We adopt a Cartesian framework, (cf. Fig.1 and consider by now only one dipolar particle in ${ 0}$), where the incident wave  has ${\bf s}_i=(0,0,1)$, and ${\bf e}_{i}=e (1,\pm i, 0)$, ${\bf b}_{i}=ne (\mp i,1,  0)$; $n=\sqrt{\epsilon\mu}$. The upper and lower sign stands for left (LCP) and right (RCP)  circular polarization.
Introducing Eqs. (\ref{es}) - (\ref{dipm}) into  (\ref{torque1}),  writing $ dS= r^2 \sin\theta d\theta d\phi $,   the interference between the incident and scattered fields in the first two terms of  (\ref{torque1}) yield integrands: $[({\bf p}^{*}\times {\bf s})({\bf e}_{i}\cdot{\bf s})+2({\bf p}\cdot {\bf s})({\bf s}\times {\bf e}_{i}^{*})+({\bf m}^{*}\times {\bf s})({\bf b}_{i}\cdot{\bf s})+2({\bf m}\cdot {\bf s})({\bf s}\times {\bf b}_{i}^{*})]$ which come from the near-zone scattered fields in the products:  $r^{3}[\epsilon({\bf s}\times {\bf E}^{(s) *})({\bf E}^{(i)}\cdot{\bf s})+\epsilon({\bf s}\times {\bf E}^{(i) *})({\bf E}^{(s)}\cdot{\bf s})+\mu^{-1}({\bf s}\times {\bf B}^{(s) *})({\bf B}^{(i)}\cdot{\bf s})+\mu^{-1}({\bf s}\times {\bf B}^{(i) *})({\bf B}^{(s)}\cdot{\bf s})]$. 
Notice that these $r^{-3}$  near-zone scattered fields [cf. Eqs. (\ref{es}) and (\ref{bs})] are those for which there is cancellation of the $r^{3}$ factor in the integrand of (\ref{torque1}), and hence lead to results independent of  the radius of integration $r$. By contrast, the contribution of  far and middle-zone terms give results proportional to $r^2$ and $r$, , which are zero allowing the integration sphere  to be arbitrarily close to the enclosed particle, which is treated like a point. Thus the integrals over the polar and azimuthal angles lead   to 
\be
<{\bf \Gamma}_{0}>=
\frac{1}{2  } \Re [ ({\bf p}\times {\bf E}^{(i) *} ) + ( {\bf m}\times{\bf B}^{(i) *} )] .  \label{torq2}
\ee
 $<{\bf \Gamma}_{0}>$ arising from the  interference of the incident and near-zone scattered fields  (in particular, when extremely close to the object they are static) comes from the extinction of the incident angular momentum, (in fact of the spin part, SAM),  and its transference to the particle. For an incident plane wave there is no contribution of the orbital angular momentum (OAM) to this {\it extinction optical torque} (EOT) $<{\bf \Gamma}_{0}>$.

On  the other hand, the interference of the scattered fields in the integrand of (\ref{torque1}) contain  terms: $2[({\bf p}\cdot {\bf s})({\bf p}^{*}\times {\bf s})+({\bf m}\cdot {\bf s})({\bf m}^{*}\times {\bf s})]$  which arise from the product of the middle-zone with the far-zone  fields with an $r$-dependence: $r^{-2}$ and $r^{-1}$, respectively,  [cf. Eqs. (\ref{es}) - (\ref{bs}) or (\ref{dipe}) - (\ref{dipm})]: $r^{3}[\epsilon({\bf s}\times {\bf E}^{(s) *})({\bf E}^{(s)}\cdot{\bf s})+\mu^{-1}({\bf s}\times {\bf B}^{(s) *})({\bf B}^{(s)}\cdot{\bf s})$, again cancelling the $r^3$ factor of the integrand.  After angular integration, they lead to the {\it scattering}, or {\it recoil}, contribution to the optical torque:
\be
<{\bf \Gamma}^{(s)}>=-\frac{k^3}{3}  [\frac{1}{\epsilon}\Im( {\bf p}^{*}  \times {\bf p}) + \mu \Im ( {\bf m}^{*}  \times {\bf m}) ]. \label{ttss1}
\ee
$\Im$ means imaginary part. As said,  $<{\bf \Gamma}^{(s)}>$  equals   the rate of radiation of electromagnetic  angular momentum to infinity $<d{\bf J}/dt>$ from the oscillating electric and magnetic dipoles \cite{jackson}. This is logical on taking into account that these dipoles are induced on the particle by the incident wave and then re-radiate. To this component, half is contributed by the OAM and the other half by the SAM.

Therefore, in summary, (\ref{torque1}), (\ref {torq2}) and (\ref{ttss1}) yield the total time-averaged torque on the particle due to an incident plane wave  as
\be
<{\bf \Gamma}>=<{\bf \Gamma}_{0}>+<{\bf \Gamma}^{(s)}>. \label{torque}
\ee
Which contains the contribution of the scattered, or recoil, component $<{\bf \Gamma}^{(s)}>$,   Eq.(\ref{ttss1}),  added to $<{\bf \Gamma}_{0}>$,   Eq.(\ref{torq2}). The former, like the latter, is an intrinsic torque. 

Eq.  (\ref{torque}) is the first main contribution of this letter.

In this regard it is interesting that the ROT (\ref{ttss1}) keeps a formal analogy with the electric-magnetic component [given by $\Re ( {\bf p}\times {\bf m}^{*})$]  of the electromagnetic force on a magnetodielectric dipolar particle, as it also arises by interference of the scattered fields  \cite{MNV2010}. On the other hand, the term $<{\bf \Gamma}_{0}>$, Eq. (\ref {torq2}) ,  is formally analogous to  the time-averaged Lorentz's force on an electric  (magnetic) dipole from a time-varying  magnetic (electric) field, given in terms of  $\Im [ {\bf p}\times {\bf  b}_{i}^{*} ]$,  ($ \Im [ {\bf m}\times{\bf e}_{i}^{*}] $).  Thus the RT plays a role in the OT as important as the  $e-m$ electric-magnetic interaction force on an electric-magnetic dipole system. As a matter of fact, like the omission of the $e-m$ interaction avoids the prediction of phenomena like Kerker's effects or pulling forces \cite{chen,novits, dogariu1}, that of the RT prevents the obtention of  a NOT, i.e. a torque on the body counterclockwise with respect to the helicity of the incident light.

Let us assume the particle being chiral, then $\alpha_{em}=-\alpha_{me}$. Introducing now Eqs. (\ref{consti}) into  (\ref{torq2}) - (\ref{torque}) and using the optical theorem, which expressing the conservation of energy  imposes  a well-known condition (cf. Eq.(16) of \cite{MNV2010})
between polarizabilities, which for a chiral particle generalizes to:
\be
\epsilon\frac{\sigma^{(a)}}{4\pi k}+\frac{2 k^3}{3 }\{\epsilon^{-1} |\alpha_{e}|^{2}   + n^2 \mu |\alpha_{m}|^2 + 2\mu |\alpha_{me}|^{2} \mp  \,\,\,\,\,\,\,\,\,\,\,\,\,\, \nonumber \\
 2n\Im [ \epsilon^{-1} \alpha_{e}^{*}   \alpha_{me} 
+ \mu \alpha_{me} \alpha_{m}^{*}]
\nonumber \\
=\pm 2n\alpha_{me}^{R}+ (\alpha_{e}^{I} +n^2  \alpha_{m}^{I})   . \,\,\,\,\,\,\,\,\,\,\,\label{top2}
\ee
($\sigma^{(a)}$
representing the particle absorption cross-section),  we write the torque on the chiral paticle due to a CP incident plane wave, as:
\be
<{\bf \Gamma}>=\pm {\bf \hat z}e^2 \epsilon\frac{\sigma^{(a)}}{4\pi k} \label{torqabs}
\ee
where ${\bf \hat z}$ is the unit vector along $OZ$ and  the signs plus and minus correspond to LCP and RCP, respectively.

Hence the OT on one bi-isotropic  (or in particular, chiral) body illuminated by a plane wave is essentially measured by its absorption cross-section. This  coincides with  previous knowledge derived both from  Mie theory and experiments \cite{laporta,mars,dunlop} in  large purely dielectric isotropic spheres. Therefore, (\ref{torqabs}) shows that even if the particle is magnetodielectric and bi-isotropic, being dipolar, it experiences an OT due to a plane wave  which is essentially given by the   absorption cross section, thus the OT being zero if the particle is non absorbing. Unlike the OF, the OT does not change depending on whether the particle satisfies Kerker's  \cite {kerk, suppre} or any other magnetodielectric conditions. Only the absorption cross section plays a role. As seen, this characteristic is  a consequence of the conservation of both energy, ruling Eq.(\ref{top2}), and angular momentum, leading to (\ref{torque1}).  This is the second contribution of this letter. 

The aforementioned duality of the OT torque and OF equations on dipolar particles is even more notorious when the {\it  incident field is an   arbitrary wave}. To do this, it is convenient to represent the fields, both in the near and intermediate regions, by the  angular spectrum of plane waves, \cite{nietolib,mandel}.



In this case a calculation with  (\ref{torque1}) shows that Eq. (\ref{torque})  applies with the ROT $<{\bf \Gamma}^{(s)}>$  remaining as in (\ref{ttss1}), however  the EOT  now becomes instead of  $<{\bf \Gamma}_0>$:

\be
 <{\bf \Gamma}^{(0)}>= <{\bf \Gamma}_0> \,\,\,\,\,\,\,\,\,\,\,  \nonumber \\
+\frac{3}{4k} \sqrt{\frac{\epsilon}{\mu}}\Im\{\frac{1}{\epsilon}({\bf p} \cdot\nabla){\bf B}^{(i) *}-\mu  ({\bf m} \cdot\nabla){\bf E}^{(i) *} \} .\,\,\,\,\,\,\,\,\,\,\,  \label{torq3}
\ee
 Where now the contribution from the OAM  to the second term of (\ref{torq3})  is half that from the SAM, and $<{\bf \Gamma}_0>$, contributed by the SAM, is given by Eq.(\ref{torq2}) with the vectors ${\bf E}^{(i)}$ and ${\bf B}^{(i)}$ of course being those of the arbitrary incident field. Notice that the additional terms appearing in (\ref{torq3}) are the formally analogous to those of the "dipolar force" which in the OF add to the "Lorentz's force terms, and are due to the spatial structure of the incident field. This is further seen by writing (\ref{torq3})  as
\be
 <{\bf \Gamma}^{(0)}>=- \frac{<{\bf \Gamma}_0>}{2}  \,\,\,\,\,\,\,\,\,\,\,\ \nonumber \\
 +\frac{3}{4k} \sqrt{\frac{\epsilon}{\mu}}\Im\{
\frac{1}{\epsilon}p_j \partial_i B_{j}^{(i) *}- \mu m_j \partial_i E_{j}^{(i) *}\}. \,\,\,\,\,\,\,\,\,\,\,\,\label{lprimfin}
\ee
Eqs. (\ref{torq3}) and (\ref{lprimfin}) are the third main contribution of this letter. 

Eq.(\ref{lprimfin}) has a formal analogy with the well-known expression  for the extinction  OF (i.e. that containing the {\it electric} and the {\it magnetic} OF) on a magnetodielectric dipolar object \cite{MNV2010}. Indeed, adding (\ref{lprimfin}) and  (\ref{ttss1})  in (\ref{torque}), the  analogy of  the OT and the OF on a magnetodielectric dipolar particle is complete as long as the duality of factors ${\bf p}$ and ${\bf m}$, and  ${\bf E}^{(i)}$ and ${\bf B}^{(i)}$, appearing in the OT and in the OF is concerned. It should be stressed that now $<{\bf \Gamma}>$ is no longer proportional to $\sigma^{(a)}$.

Next, we  illustrate this analysis addressing the phenomenon of NOT  in sets of  dipolar particles.  We consider two identical  dielectric spheres with centers at ${\bf r}_0=(0,0,0)$ and ${\bf r}_0=(0,R,0)$, (cf. Fig. 1). Illumination occurs with a plane CP wave. Considering the coupling between the dipoles induced in both particles, the resulting OT on the two sphere system is:
\be
<{\bf \Gamma}>= \pm {\bf \hat z} e^2 |\alpha_e|^2 {\cal R}\{(\frac{2}{3}k^3|\alpha_e|^2 + \epsilon\frac{\sigma^{(a)}}{4\pi k}){\cal R}+\alpha_{e}^{R }{\cal I}    \}. \,\,\,\, \,\,\, \label{torque9}
\ee
The upper and lower sign of $\pm$ applying to LCP and RCP, respectively. $R$ and $I$ denote real and imaginary parts.

Or, by using (\ref{top2})
\be
<{\bf \Gamma}>= \pm {\bf \hat z} e^2 |\alpha_e|^2 {\cal R}\{\alpha_{e}^{I}{\cal R}+\alpha_{e}^{R} {\cal I}    \}. \label{torque10}
\ee
Where
\be
\eta_{\parallel}=\frac{2 e^{ikR}}{\epsilon R^3}(1-ikR);  \,\, \eta_{\perp} =\frac{2 e^{ikR}}{\epsilon R^3} (k^2 R^2 +ikR-1). \,\,\,\, \,\,\,\label{paref}
\ee
are  the tensor Green function eigenvalues, so that the real and imaginary parts of $\eta_{\parallel} - \eta_{\perp}$ are ${\cal R}=\Re[\eta_{\parallel} - \eta_{\perp}] =\frac{k^3}{3\epsilon}kR f_{\parallel}(kR) 
+\frac{1}{\epsilon R^3}(2kR \sin kR+3 \cos kR)$. 
${\cal I}=\Im[\eta_{\parallel} - \eta_{\perp}] =\frac{2k^3}{3\epsilon}[f_{\parallel}(kR)-f_{\perp}(kR)]-\frac{2kR \cos kR}{\epsilon R^3}$. 
\begin{figure}[htbp]
\centerline{\includegraphics[width=.8\columnwidth]{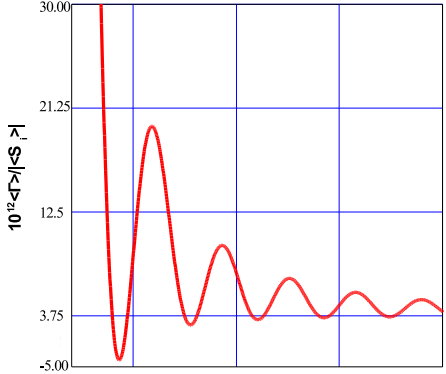}}
\caption{(Color online). Averaged optical torque, normalized to incident energy flow $<|{\bf S}_{i}|>$, versus $kR$ on two Si particles ($n_p=3.5$) separated a distance  $R$. ($\alpha_{e}^{R}=7.91 \times 10^{6} nm^3$,  
 $\alpha_{e}^{I}=1.1 \times  10^7 nm^3$, $c=3\times 10^{17} nm\cdot s^{-1}$). The particles rotate with respect to an axis parallel to the $OZ$ direction, at  $y=R/2$. There are regions where $<{\bf \Gamma}>$ is negative.}
\end{figure}

 It is curious that $f_{\parallel}(\chi)$ and  $ f_{\perp}(\chi)$ are the parts (parallel and perpendicular, respectively, to the the $OY$ segment joining both particles)  of the diagonal elements  of the correlation tensor  $f_{ij}(\chi)$   ($i,j=1, 2, 3=x,y,z$) of  an statistically homogeneous and isotropic random field ${\bf E}^{r}({\bf r})$,  namely $f_{ii}(\chi)=<{E}_{i}^{r}({\bf r}){ E}_{i}^{r}({\bf r}')>/<|{E}_{i}^{r}({\bf r})|^2>$, $\chi=|{\bf r}-{\bf r}'|$.  So that $f_{\parallel}(\chi)=f_{yy}(\chi)$ and $f_{\perp}(\chi)=f_{xx}(\chi)=f_{zz}(\chi)$ \cite{eck}:
$f_{\parallel}(\chi)=3( sin \chi-\chi cos \chi)/\chi^3$; 
$f_{\perp}(\chi)=(3/2)[  sin \chi/\chi-(sin \chi-\chi  cos \chi)/\chi^3]$.
Obviously, when the particles approach each other, one has ${\cal R}\rightarrow  3/ \epsilon R^3$,  ${\cal I}\rightarrow 0$,  ($ kR\rightarrow 0$). 

One may wonder on the appearence of $f_{ij}(\chi)$   in this theory, however, this is not surprising as these functions are associated to  $\eta_{\parallel}$ and $\eta_{\perp}$, and appear in any interaction  with a $kR$ dependence characterized by coupled dipoles. Notice, on the other hand, the oscillating behavior of the OT exerted on the two spheres versus $kR$, according to Eqs.(\ref{torque9}) or (\ref{torque10}); this encompasses the existence of a NOT on the system.

As an example, we consider a LCP plane wave  incident on two Si spheres of radius $230 nm$ in air  at $\lambda \simeq 1350 nm$, where the electric dipole dominates \cite{Nieto2011}. As seen there are separations $R$ leading to a negative OT.  $\alpha_{e}^{R}$ influences mucho more than $\alpha_{e}^{I}$ the results in (\ref {torque10}). This torque  (cf. Fig. 2) is enhanced in the electric dipole  resonance region in this kind of high $n_p$ particles since, on comparison, we obtained a torque  one order of magnitude smaller in the equivalent region ($\lambda=700 nm$) if being of the same size, they are both of latex, ($n_p=1.5$).

In conclusion, the expression of the ROT   is  general  for any illuminating field. Our generalized EOT, together with the ROT, may give rise for  specially  structured incident beams, to negative torques on one dipolar object.  Further studies for other sets of magnetodielectric, bi-isotropic, coupled dipoles remain to be done.

Work  supported by the MINECO through grants FIS2012-36113-C03-03 and FIS2014-55563-REDC. The author appreciates discussions with M.V. Berry and J.J. Saenz that triggered his interest on this subject.

\end{document}